\documentclass[aps,prl,superscriptaddress,twocolumn,amsmath,amssymb]{revtex4-1} 

\usepackage{graphicx}
\usepackage{bm}
\usepackage{epstopdf}

\begin{document}

\newcommand{\etal}{{\it et al.}\/}
\newcommand{\gtwid}{\mathrel{\raise.3ex\hbox{$>$\kern-.75em\lower1ex\hbox{$\sim$}}}}
\newcommand{\ltwid}{\mathrel{\raise.3ex\hbox{$<$\kern-.75em\lower1ex\hbox{$\sim$}}}}

\title{The overdoped end of the cuprate phase diagram}

\author{Thomas A. Maier}
\affiliation{Computational Sciences and Engineering Division, Oak Ridge National Laboratory, Oak Ridge, Tennessee 37831-6164, USA}
\affiliation{Center for Nanophase Materials Sciences, Oak Ridge National Laboratory, Oak Ridge, Tennessee 37831-6164, USA}
\author{Seher Karakuzu}
\affiliation{Center for Nanophase Materials Sciences, Oak Ridge National Laboratory, Oak Ridge, Tennessee 37831-6164, USA}

\author{Douglas J.~Scalapino}
\affiliation{Department of Physics, University of California, Santa Barbara, CA 93106-9530, USA}

\date{\today}

\begin{abstract}
Studying the disappearance of superconductivity at the end of the overdoped region of the cuprate phase diagram offers a different approach for investigating the interaction which is responsible for pairing in these materials. In the underdoped region this question is complicated by the presence of charge and stripe ordered phases as well as the pseudogap. In the overdoped region the situation appears simpler with only a normal phase, a superconducting phase and impurity scattering. Here, for the overdoped region, we report the results of a combined dynamic cluster approximation (DCA) and a weak Born impurity scattering  calculation for a $t-t'-U$ Hubbard model. We find that a decrease in the $d$-wave pairing strength of the two-particle scattering vertex is closely coupled to changes in the momentum and frequency structure of the magnetic spin fluctuations as the system is overdoped. Treating the impurity scattering within a disordered BCS $d$-wave approximation, we see how the combined effects of the decreasing $d$-wave pairing strength and weak impurity scattering  lead to the end of the $T_c$ dome.
\end{abstract}


\maketitle

\section*{Introduction}

In the overdoped region of the cuprate phase diagram the normal phase exhibits properties similar to those of a strongly correlated Fermi liquid \cite{Proust,Bangura,Kramer}. The pseudogap as well as the charge and stripe ordered phases, which compete or coexist with superconductivity at lower doping, are absent \cite{Doiron-Leyraud}. This is a region in which the results of numerical calculations are expected to be less sensitive to boundary conditions and lattice size effects. Here, using a dynamic cluster approximation (DCA) \cite{Maier1} we study the decrease in the strength of the d-wave pairing interaction for a $t-t'-U $ Hubbard model in the overdoped regime as the density $x$ of doped holes per site increases \cite{Huang,Maier0}. Then, including impurity scattering within a disordered Bardeen-Cooper-Schrieffer (BCS) $d$-wave approximation \cite{Lee-Hone}, we examine the end of the $T_c$ dome.

The Hubbard model we will study has a near-neighbor hopping $t$, a next-near-neighbor hopping $t'$ and an onsite Coulomb interaction $U$. 
\begin{align} H =&-t\sum_{\langle i,j\rangle\sigma}c^\dagger_{i\sigma}c^{\phantom\dagger}_{j\sigma}-
	t'\sum_{\langle\langle i,j\rangle\rangle\sigma}c^\dagger_{i\sigma}c^{\phantom\dagger}_{j\sigma}\nonumber\\
	&+U\sum_in_{i\uparrow}n_{i\downarrow}-\mu\sum_{i\sigma}n_{i\sigma}
\label{eq:1}
\end{align}
The tight binding parameters give rise to a bandstructure $\varepsilon_{\bm k}=-2t(\cos k_x+\cos k_y)-4t'\cos k_x\cos k_y$ and $\mu$ controls the filling, which we will measure in terms of the density of holes $x$ away from half-filling. In the following, the results for $t'/t=-0.25$ and $U/t=7.0$ were obtained from a dynamic cluster approximation using a 12-site cluster (Fig.~\ref{fig:1} in Ref.~\cite{Maier1}) and a continuous-time auxiliary field quantum Monte Carlo algorithm to solve the DCA cluster problem.

\section*{Results}

\begin{figure}[htbp]
 \includegraphics[width=0.35\textwidth]{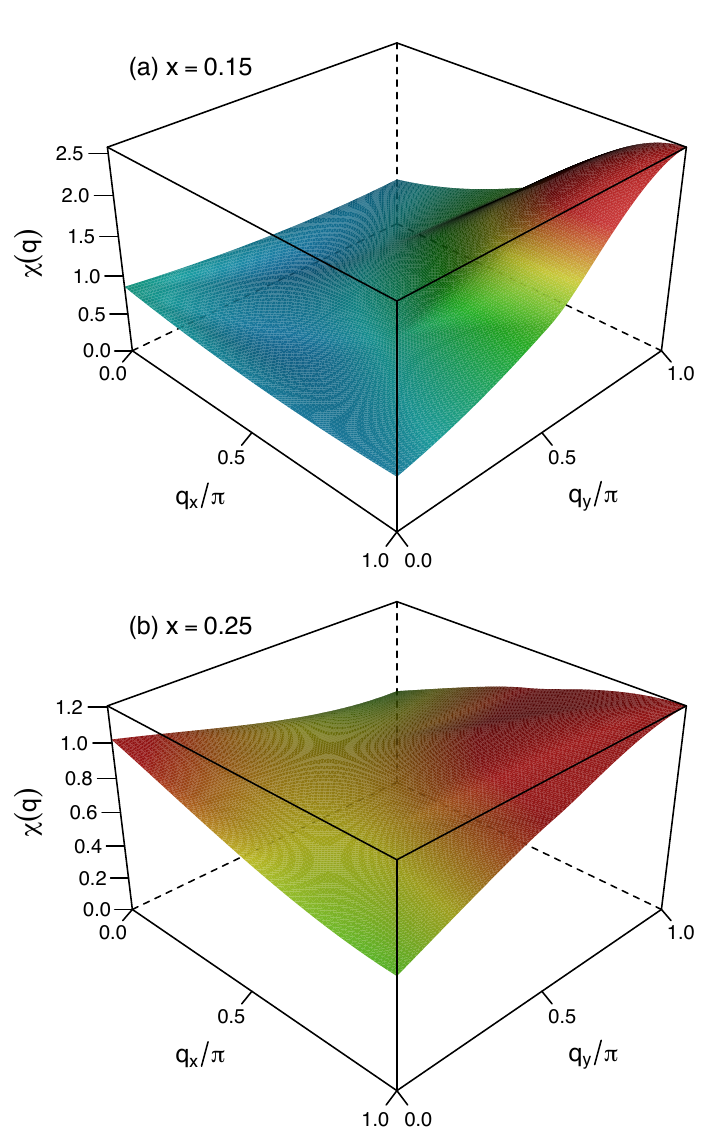}
  \caption{The spin susceptibility $\chi({\bm q})$ at $T=0.1t$ for (a) $x=0.15$ and (b) 0.25. As the doping increases the antiferromagnetic response weakens and there is an increase in the ferromagnetic response.\label{fig:1}}
\end{figure}

As  discussed in the Supplemental Material section \cite{ref:SM}, previous DCA calculations  \cite{Chen,Wu} have found that for these parameters the pseudogap ends for $ x \gtrsim 0.15$. This is the overdoped regime that we will study. In Fig.~\ref{fig:1} we have plotted the spin susceptibility $\chi({\bm q})$ at $T=0.1t$ for dopings $x=0.15$ and $x=0.25$.

\begin{figure}[htbp]

\includegraphics[width=0.42\textwidth]{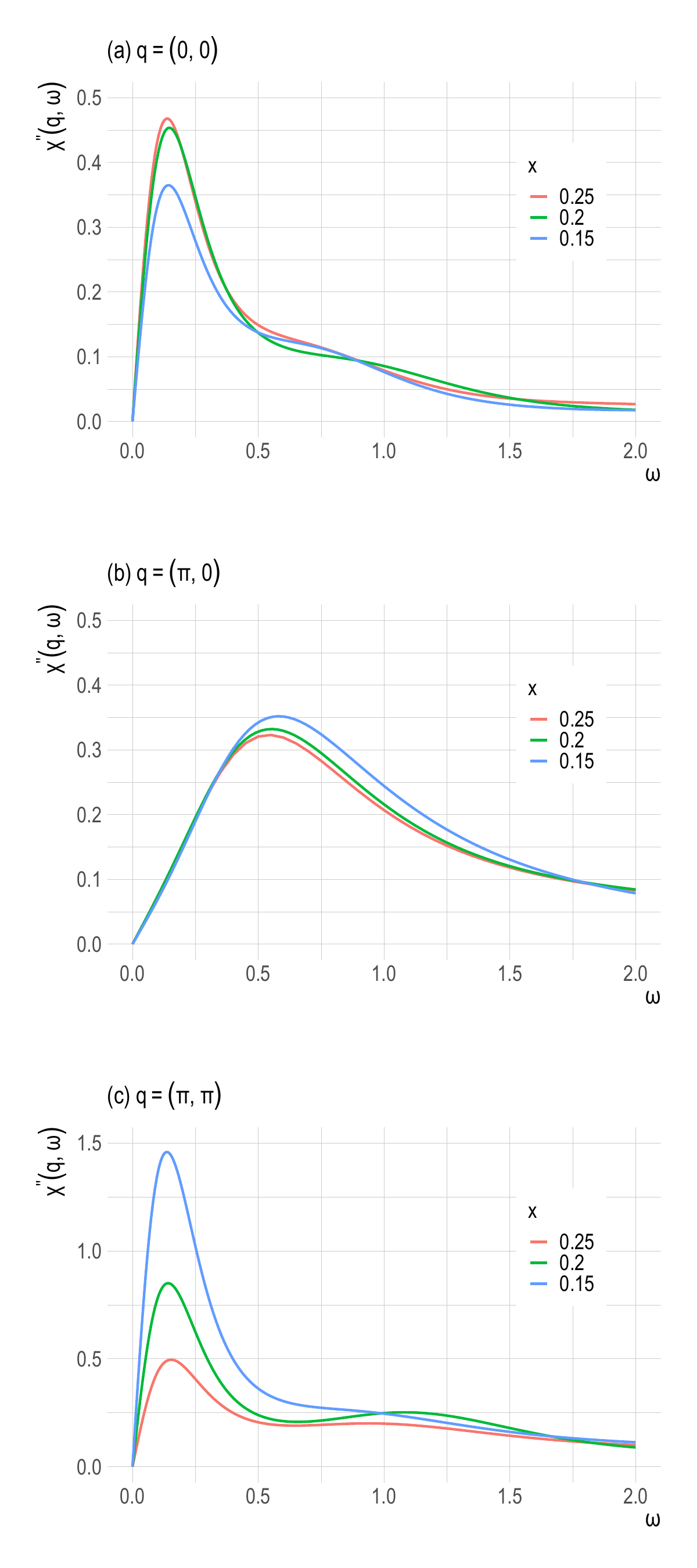}
  \caption{The imaginary part of the DCA cluster spin susceptibility $\chi''({\bm q},\omega)$ vs.\ $\omega$ for various cluster momenta and dopings $x=0.15$ (blue), 0.2 (green), and 0.25 (red). The spin fluctuation spectral weight at large (AF) momenta is reduced as the doping increases while the long wavelength (FM) spectral weight increases. For intermediate momenta ${\bm q}=(\pi,0)$, the high energy spin-fluctuation spectral weight remains. \label{fig:2}} 
  \end{figure}

As the doping  increases, the susceptibility at large  momentum transfer ${\bm q}=(\pi,\pi)$ is reduced and the ferromagnetic (FM) spin susceptibility at small momentum transfer is increased. However, as previously noted \cite{Huang}, a significant response also remains at intermediate values of momentum transfer ${\bm q}=(\pi/2,\pi/2)$ and ${\bm q}=(\pi,0)$. Further insight into the evolution of the spin-fluctuations with doping is illustrated in Fig.~\ref{fig:2} where we have plotted the spin-fluctuation 
spectral weight $-\frac{1}{\pi}{\rm Im}\chi({\bm q},\omega)$ at different momentum transfers for various dopings. These results were obtained from a Maximum Entropy analytic continuation of the DCA imaginary time data. At large momentum transfer ${\bm q}=(\pi,\pi)$, one sees that the spin-fluctuation spectral weight is significantly reduced with doping, while at intermediate values of momentum transfers spin-fluctuations remain. For example, at the Brillouin zone boundary ${\bm q}=(\pi,0)$, the magnetic excitations are relatively unchanged with doping. This type of behavior has been seen in inelastic neutron scattering (INS) and resonant inelastic X-ray scattering (RIXS)  experiments \cite{Roberts,Ament,Wakimoto,Meyers}.

Next we turn to the strength of the pairing interaction in the overdoped region. The particle-particle Bethe-Salpeter equation is
\begin{equation} \lambda_\alpha\phi_\alpha(k)=-\frac{T}{N}\sum_{k'}
	\Gamma(k,k')G(k')G(-k')
	\phi_\alpha(k')\label{eq:2}
\end{equation}
Here $G$ is the single particle propagator and $\Gamma$ is the irreducible particle-particle scattering vertex and we have used $k=({\bm k},\omega_n)$. At the superconducting transition temperature $T_c$ the leading eigenvalue of Eq.~(\ref{eq:2}) goes to 1. For the doped Hubbard model the eigenfunction $\phi_d({\bm k},\omega_n)$ with the leading eigenvalue has $d$-wave symmetry. At a doping $x=0.15$ the DCA calculations give $\lambda_d(T_c)=1$ with $T_c/t=0.027$. For a near-neighbor hopping $t=0.2$ eV, this would correspond to a transition temperature $T_c\sim 65$ K. Here we are interested in what happens to the strength of the pairing interaction as $x$ increases and the system is overdoped. 

Multiplying Eq.~(\ref{eq:2}) by $\phi_d({\bm k},\omega_n)$ and summing over $({\bm k},\omega_n)$ one obtains the following expression for $\lambda_d$ 
\begin{equation}
  \lambda_d=\frac{-\frac{T^2}{N^2}\sum_{k,k'}\phi_d(k)
	\Gamma(k,k')G(k')G(-k')
	\phi_d(k')}{\frac{T}{N}\sum_{k}\phi^2_d(k)}\label{eq:3}
\end{equation}
Then inserting a complete set of states between the vertex $\Gamma$ and the $GG$ propagators and assuming that the leading $d$-wave eigenvalue is dominant, one obtains  the separable approximation 
\begin{equation}
  \lambda_d(T)\simeq V_d(T)P_{d0}(T)\label{eq:4}
\end{equation}
with the strength of the pairing interaction 
\begin{equation}
  V_d(T)=-\frac{\sum_{k}\sum_{k'}\phi_d(k)
	\Gamma(k,k')\phi_d(k')}{\left(\sum_{k}
	\phi^2_d(k)\right)^2}\label{eq:5}
\end{equation}
and the non-interacting but dressed  two-particle pairfield susceptibility
\begin{equation}
  P_{d0}(T)=\frac{T}{N}\sum_{k}\phi^2_d(k)G(k)G(-k)
	\label{eq:6}
\end{equation}
In evaluating these expressions we will approximate the $d$-wave eigenfunction
\begin{equation}
  \phi_d({\bm k},\omega_n)\sim(\cos k_x-\cos k_y)
	\frac{(\pi T)^2+\omega^2_c}{\omega^2_n+\omega^2_c}\label{eq:7}
\end{equation}
with $\omega_c=t$. This form provides a reasonable approximation and is less noisy than using the DCA eigenfunction $\phi({\bm k},\omega_n)$.

\begin{figure}[htbp]
\includegraphics[width=0.5\textwidth]{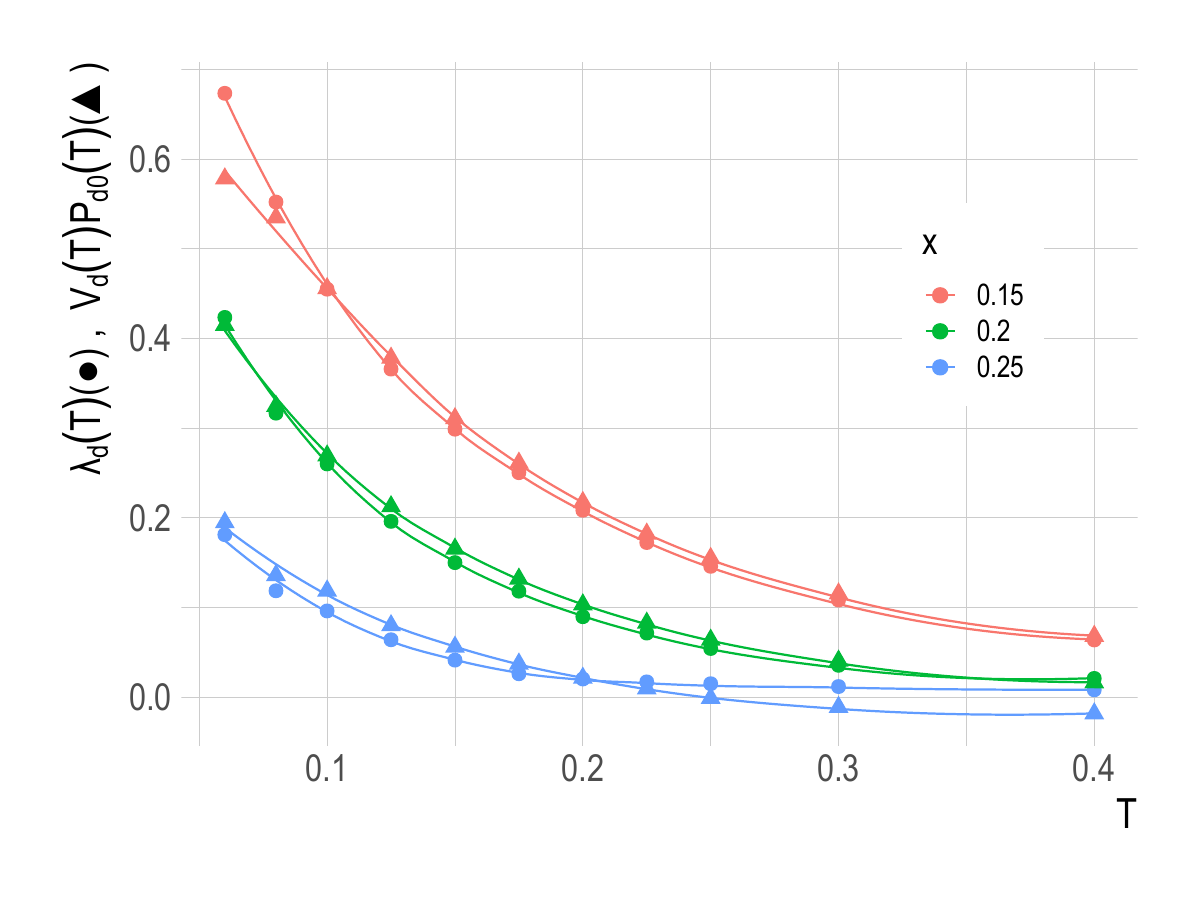}
  \caption{A comparison of the $d$-wave eigenvalue $\lambda_d(T)$ (\textbullet) of the Bethe-Salpeter equation~(\ref{eq:2}) with the separable approximation $P_{d0}(T)V_d(T)$ ($\blacktriangle$) given by Eqs.~(\ref{eq:5}) and (\ref{eq:6}) for several values of the doping $x$. The quality of the fit reflects the dominance of the leading $d$-wave eigenvalue. \label{fig:3}}
\end{figure}

\begin{figure}[htbp]
 \includegraphics[width=0.5\textwidth]{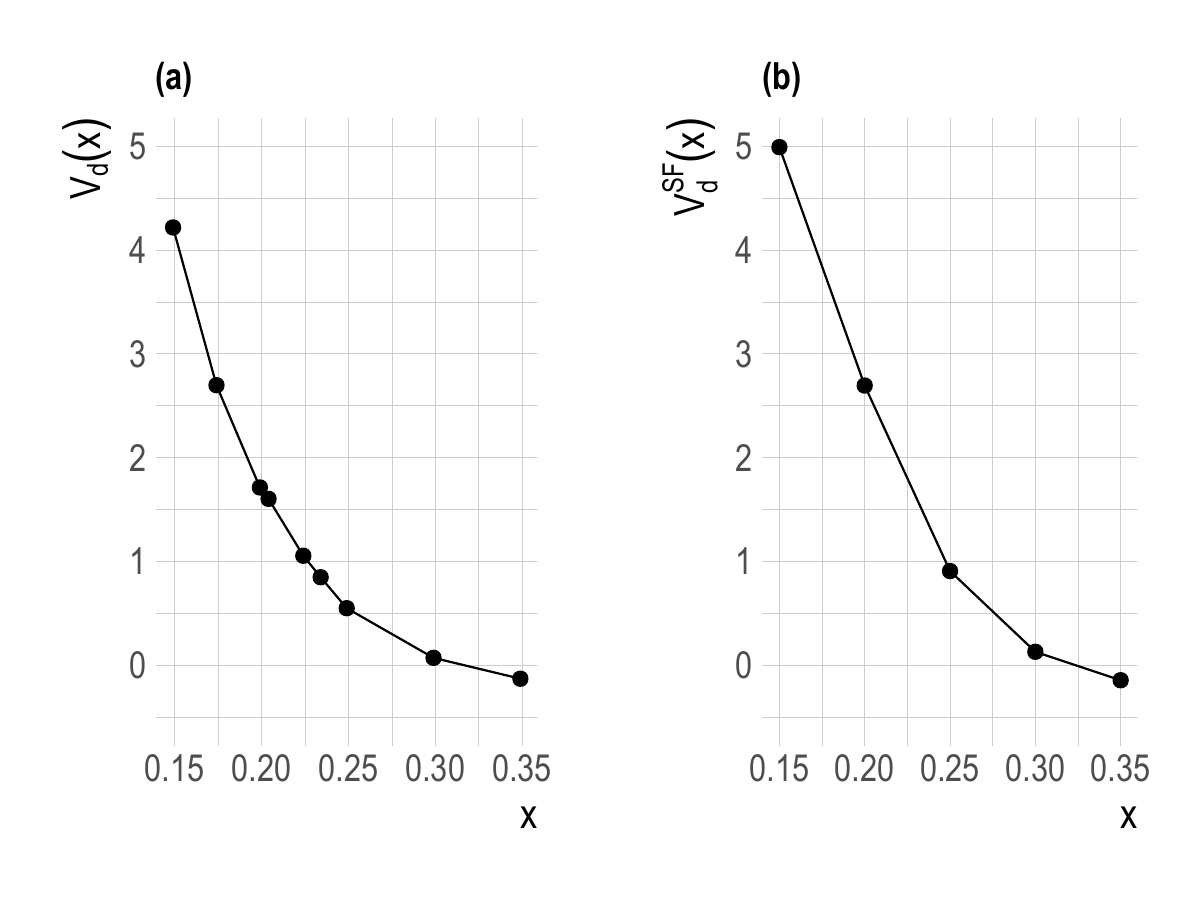}

  \caption{	a) The $d$-wave pairing strength $V_d(x)$, Eq.~(\ref{eq:5}), at $T=0.08t$ obtained from the 2-particle scattering vertex $\Gamma$ plotted versus the doping $x$. (b) The $d$-wave spin-fluctuation pairing strength, $V^{\rm SF}_d(x)$ Eq.~(\ref{eq:8}), at $T=0.08t$ versus $x$. \label{fig:4} }
\end{figure}

To check the validity of the separable approximation for  $\lambda_d(T)$ given by Eq.~(\ref{eq:4}) we have plotted $\lambda_d(T)$ and the product $V_d(T)P_{d0}(T)$ versus $T$ in Fig.~\ref{fig:3} for different dopings. The close agreement between $\lambda_d(T)$ and $V_d(T)P_{d0}(T)$ seen in 
Fig.~\ref{fig:3} arises from the fact that, while there are other singlet eigenstates of the Bethe-Salpeter equation such as extended $s$-wave and odd frequency $p$-waves, the singlet channel is dominated by the $d$-wave eigenfunction. Thus $V_d$ provides a measure of the $d$-wave pairing strength associated with the two particle scattering vertex $\Gamma({\bm k},\omega_n,{\bm k}',\omega_{n'})$. Results for $V_d(x)$ at a low temperature $T=0.08t$ are shown in Fig.~\ref{fig:4}a. Here one sees the decrease of the $d$-wave coupling strength as the hole doping is increased.

In spin-fluctuation theories of the pairing interaction, a measure of the strength of the $d$-wave pairing interaction is given by 
\begin{equation} 
	V^{\rm SF}_d=\frac{3\bar U^2}{2}\frac{1}{N}\sum_{\bm q}\int^\infty_0\frac{d\omega}{\pi}
		\frac{{\rm Im}\chi({\bm q},\omega)}{\omega}\cos q_x\label{eq:8}
\end{equation}
Using the DCA results for the cluster spin susceptibility $\chi({\bm q},\omega)$ and replacing $\bar U$ by $U/2$, as found in previous DCA studies \cite{Maier2}, results for $V^{\rm SF}_d(x)$ versus $x$ at $T=0.08t$ are plotted in Fig.~\ref{fig:4}b. The change in $\chi({\bm q})$ shown in Fig.~\ref{fig:1} and the shift of the spin-fluctuation spectral weight with doping shown in Fig.~\ref{fig:2} are reflected in the decrease in $V^{\rm SF}_d$ as the doping increases. 

\begin{figure}[htbp]
 \includegraphics[width=0.5\textwidth]{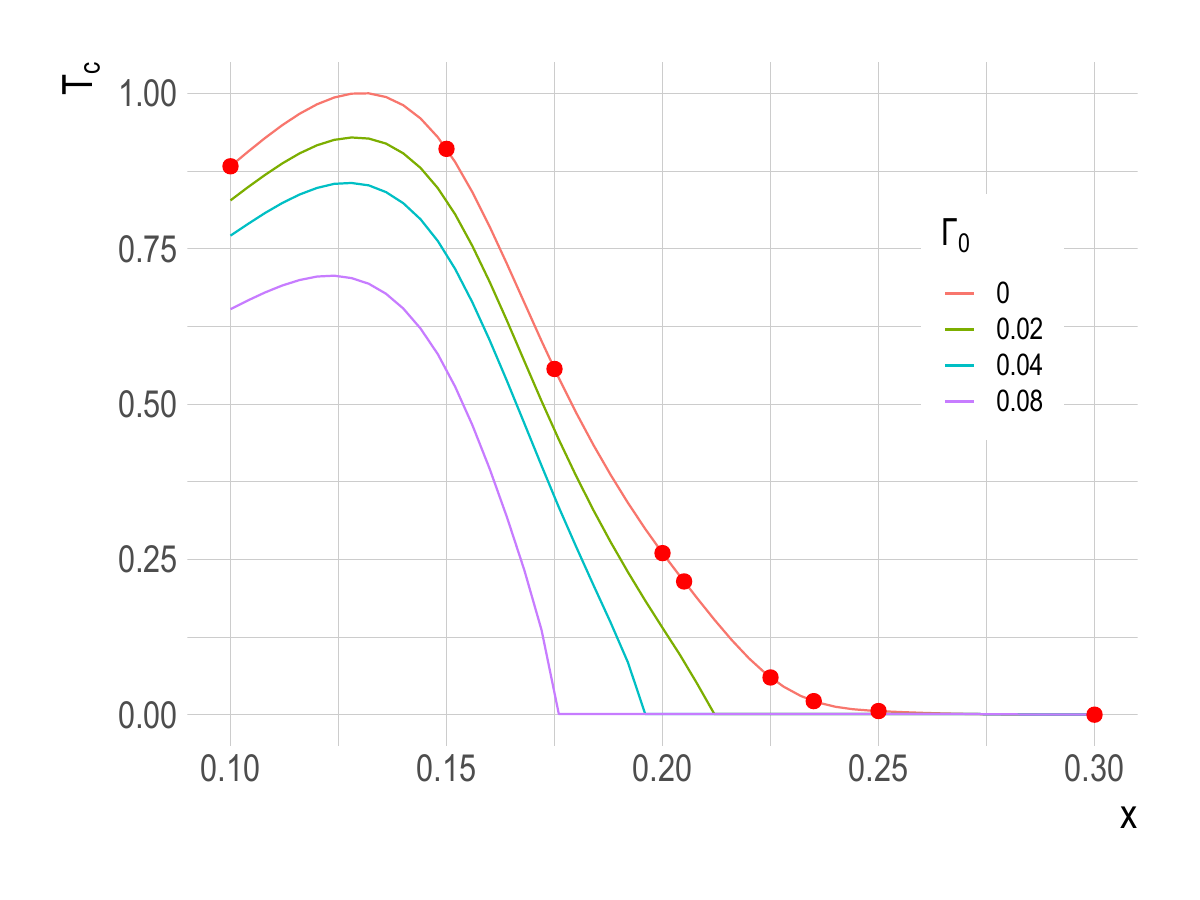}
  \caption{	The superconducting transition temperature $T_{c0}(x)$ (solid dots) for the pure system determined from an extrapolation of the Bethe-Salpeter eigenvalue $\lambda_d(T)$ to 1 for specific dopings. The red curve is a fit to these points. The additional $T_c(x)$ curves are solutions of the AG equation for different impurity scattering strengths $\Gamma_0$. The transition temperatures are normalized by the maximum $T_{c0}$ value.\label{fig:5}}
\end{figure}

The end behavior of the $T_c$ dome involves the  effects of impurities. Here we have in mind a situation in which the impurity dopants lie off of the $CuO_2$ plane, adding x holes and giving rise to weak Born impurity scattering. Within the framework of a fluctuation exchange approximation, Kudo and Yamada \cite{Kudo} found that the reduction of the Bethe-Saltpeter eigenvalue associated with the decrease in strength of the pairing interaction caused by impurity scattering is approximately off-set by the increase of the spectral weight of the single particle propagator, leaving pair breaking as the dominant effect of the impurity scattering. We will assume that this is also the case here and use the Abrikosov-Gorkov \cite{Abrikosov} expression for the superconducting transition temperature given by 
\begin{equation}\label{eq:9}
  \ln\left(\frac{T_{c0}(x)}{T_c(x)}\right)=\psi\left(\frac{1}{2}+
	\frac{\Gamma(x)}{2\pi T_c(x)}\right)-\psi(1/2)\,.
\end{equation}
Here $T_{c0}(x)$ is the putative superconducting transition temperature of the doped system without impurity scattering obtained by extrapolating the eigenvalue of the Bethe-Salpeter equation $\lambda_d(T_{c0}(x))$ to $1$. Assuming that the impurity dopants lie out of plane, $\Gamma(x)$ is the normal state Born impurity scattering rate, which we take proportional to $x$
\begin{equation}\label{eq:10}
\Gamma(x)=\Gamma_0 x 
\end{equation}
and $\psi$ is the digamma function. Results for $T_c(x)$ versus $x$ for various values of the scattering rate $\Gamma_0$ per doped hole are shown in Fig.~\ref{fig:5}.

Here $T_{c0}(x)$ vanishes with an essential singularity $\exp(-t/V_d(x))$ at an end point where $V_d(x)$ goes to zero. In the dirty $d$-wave theory, $T_c(x)$ approaches the end point $x_0$ as $(x-x_0)^{1/2}$ with $x_0$ determined by
\begin{equation}
	T_{c0}(x_0)/\Gamma(x_0) = 2\gamma/\pi\,
\end{equation}
with $\gamma\sim 1.78$.

\section*{Discussion and Conclusion}

We have used a combined DCA and weak Born impurity scattering calculation for a 2D Hubbard model to study the disappearance of superconductivity at the end of the overdoped region of the cuprate phase diagram. We have found that the decrease in the $d$-wave pairing strength with increasing doping is closely related to a similar decrease in the strength of the $d$-wave spin-fluctuation interaction. The additional effect of impurity scattering, taken into account within a disordered BCS $d$-wave approximation, is found to lead to a further reduction of $T_c$ as the doping increases. Hence, in this work, the decrease of $T_c(x)$ in the overdoped regime reflects both a decrease in the "clean" transition temperature $T_{c0}(x)$ due to a reduction in the pairing strength and an increase in the impurity scattering rate with doping. Alternatively, "dirty $d$-wave" models \cite{Lee-Hone} in which $T_{c0}(x)$ is constant and there is an increase in the Born impurity scattering, starting from a finite doping, to fit the observed $T_c(x)$ have proved very useful. However, in this case where $T_{c0}$ is a constant greater than the maximum $T_c$, the required impurity scattering rate for the dirty $d$-wave model will be considerably larger than what we have used.

\section*{Acknowledgments}

The authors would like to thank P.~J.~Hirschfeld and S.~A.~Kivelson for their helpful comments. This work was supported by the Scientific Discovery through Advanced Computing (SciDAC) program funded by the U.S. Department of Energy, Office of Science, Advanced Scientific Computing Research and Basic Energy Sciences, and Division of Materials Sciences and Engineering. An award of computer time was provided by the INCITE program. This research used resources of the Oak Ridge Leadership Computing Facility, which is a DOE Office of Science User Facility supported under Contract DE- AC05-00OR22725.


\end{document}


\newcommand{\etal}{{\it et al.}\/}
\newcommand{\gtwid}{\mathrel{\raise.3ex\hbox{$>$\kern-.75em\lower1ex\hbox{$\sim$}}}}
\newcommand{\ltwid}{\mathrel{\raise.3ex\hbox{$<$\kern-.75em\lower1ex\hbox{$\sim$}}}}

\title{The overdoped end of the cuprate phase diagram – Supplemental Material}

\author{Thomas Maier}
\affiliation{Computational Sciences and Engineering Division and Center for Nanophase Materials Sciences, Oak Ridge National Laboratory, Oak Ridge, Tennessee 37831-6164, USA}
\author{Seher Karakuzu}
\affiliation{Computational Sciences and Engineering Division and Center for Nanophase Materials Sciences, Oak Ridge National Laboratory, Oak Ridge, Tennessee 37831-6164, USA}

\author{D.J.~Scalapino}
\affiliation{Department of Physics, University of California, Santa Barbara, CA 93106-9530, USA}

\date{\today}

\maketitle

\renewcommand{\theequation}{S\arabic{equation}}
\renewcommand{\thefigure}{S\arabic{figure}}

\begin{figure}[htbp]
 \includegraphics[width=0.8\textwidth]{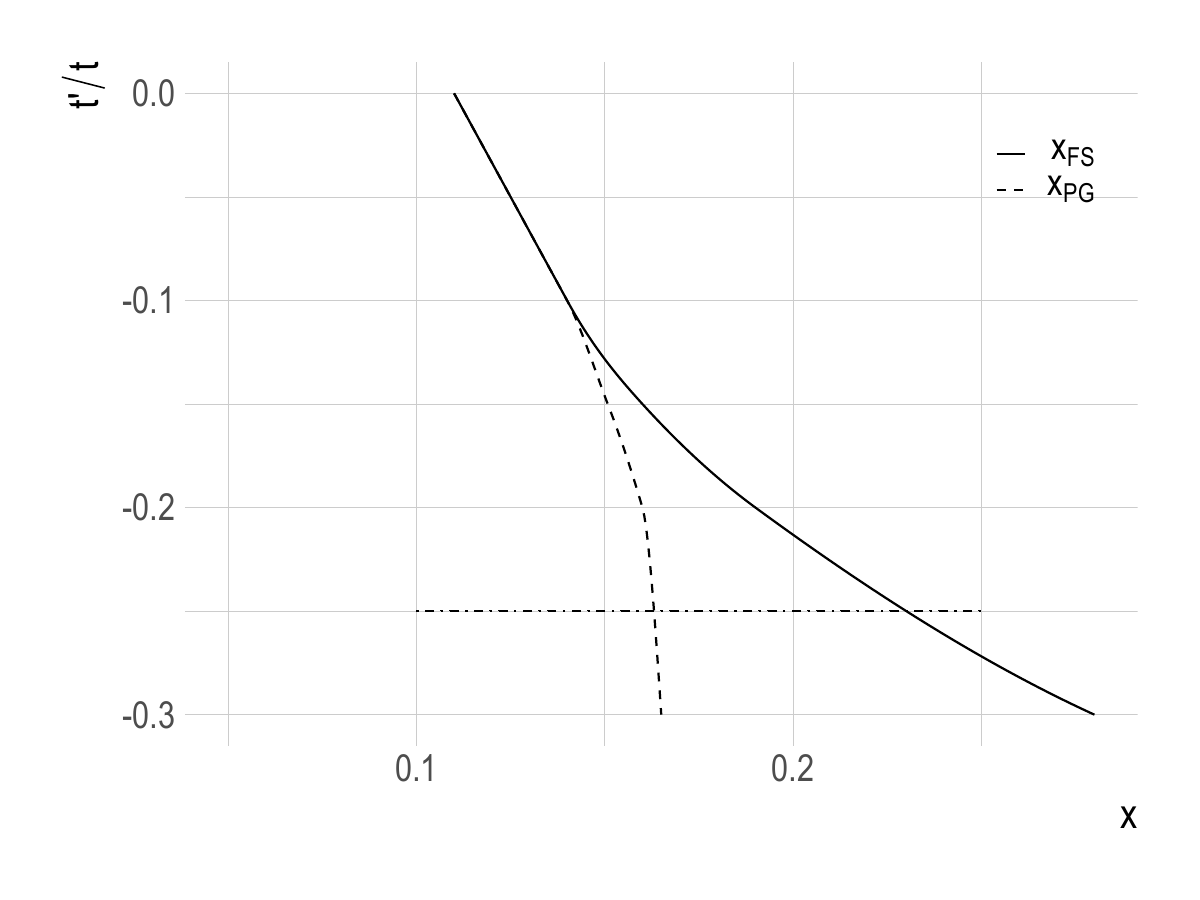}
  \caption{A part of the $t'/t-x$ phase diagram based on DCA calculations
	\protect{\cite{Wu}}. The solid curve denotes a Lifshitz transition which
	separates a lower doped region which has a hole-like FS around the $(\pi,\pi)$
	point of the Brillouin zone from a region which has an electron-like FS about
	the origin (0,0). A dashed curve, which separates from the Lifshitz curve at
	larger doping marks the end of the pseudogap region. Here we study the decrease
	of the $d$-wave pairing strength and the end of the $T_c$ dome for t'/t=-0.25 as the hole doping increases along the dash-dot line.
	\label{fig:S1}}
\end{figure}

In Fig.~\ref{fig:S1} we show a section of a $t'/t-x$ phase diagram based upon DCA
calculations by Wu \etal\ \cite{Wu}. Here a solid curve marks the Lifshitz
transition at which the topology of the Fermi surface changes from hole-like
around $(\pi,\pi)$ to electron-like around (0,0) as the hole doping increases.
The dashed curve in Fig.~\ref{fig:S1} marks the end of the pseudogap (PG) regime.
Similar to the cuprates, as discussed by Doiron-Leyraud \etal\ \cite{Doiron-Leyraud},
the simulation finds that a PG does not open on an
electron-like FS, confining the PG to a region of the $t'/t-x$ phase diagram
in which $x$ is less than the curve marking the Lifshitz doping in Fig.~\ref{fig:S1}.
However, for larger values of $|t'/t|$ there is a range of dopings below the
Lifshitz doping in which the PG is also absent \cite{Wu,Doiron-Leyraud}. In the main text, for
$t'/t=-0.25$, we examine the strength of the $d$-wave pairing and the end of the $T_c$ dome as the doping is increased along the dash-dotted line shown in Fig.~\ref{fig:S1}.